\documentclass{article}

\usepackage{algorithm,algorithmic}
 \usepackage{graphicx}
     \usepackage[preprint]{neurips_2019}

\usepackage[utf8]{inputenc} 
\usepackage[T1]{fontenc}    
\usepackage{hyperref}       
\usepackage{url}            
\usepackage{booktabs}       
\usepackage{amsfonts}       
\usepackage{nicefrac}       
\usepackage{microtype}      

\usepackage{extarrows,xfrac,bm,nameref,dsfont,color,cancel}
\usepackage{amsmath}
\newcommand{\EN}{\mathcal{EN}} 
\newcommand{\DE}{\mathcal{DE}} 
\newcommand{\G}{\mathcal{G}} 
\newcommand{\D}{\mathcal{D}}

\title{Generative Imaging and Image Processing via Generative Encoder}

\author{%
  Lin~Chen \\
  Department of Mathematics \\
  National University of Singapore \\
  10 Lower Kent Ridge Road \\
  Singapore 119076 \\
  \texttt{matche@nus.edu.sg} \\
  \And
  Haizhao~Yang \\
  Department of Mathematics and Institute of Data Science \\
  National University of Singapore \\
  10 Lower Kent Ridge Road \\
  Singapore 119076 \\
  \texttt{haizhao@nus.edu.sg} \\
}

\begin{document}

\maketitle

\begin{abstract}
 This paper introduces a novel generative encoder (GE) model for generative imaging and image processing with applications in compressed sensing and imaging, image compression, denoising, inpainting, deblurring, and super-resolution. The GE model consists of a pre-training phase and a solving phase. In the pre-training phase, we separately train two deep neural networks: a generative adversarial network (GAN) with a generator $\G$ that captures the data distribution of a given image set, and an auto-encoder (AE) network with an encoder $\EN$ that compresses images following the estimated distribution by GAN. In the solving phase, given a noisy image $x=\mathcal{P}(x^*)$, where $x^*$ is the target unknown image, $\mathcal{P}$ is an operator adding an addictive, or multiplicative, or convolutional noise, or equivalently given such an image $x$ in the compressed domain, i.e., given $m=\EN(x)$, we solve the optimization problem
 \[
 z^*=\underset{z}{\mathrm{argmin}} \|\EN(\G(z))-m\|_2^2+\lambda\|z\|_2^2
 \]
to recover the image $x^*$ in a generative way via $\hat{x}:=\G(z^*)\approx x^*$, where $\lambda>0$ is a hyperparameter. The GE model unifies the generative capacity of GANs and the stability of AEs in an optimization framework above instead of stacking GANs and AEs into a single network or combining their loss functions into one as in existing literature. Numerical experiments show that the proposed model outperforms several state-of-the-art algorithms.
\end{abstract}

\section{Introduction}

\label{sec:introduction}

Recently, deep learning-based structures have become an effective tool for imaging and image processing with compressed or compressible information. One stream of such studies is an end-to-end training with a deep neural network (DNN) mapping a source image to a reconstructed image with desired properties. Due to the powerful representation capacity of DNNs, DNNs can approximate the desired imaging or image processing procedure well as long as training data are sufficiently good. To ease the training of DNNs, special NN structures are proposed, e.g., DNNs that mimic a traditional optimization algorithms for imaging or image processing (\cite{ABEZ2016,YSLX2016,Ledig2017,L2017}), autoencoders (AEs) with built-in image compression and reconstruction (\cite{V2010,SCH2017}). The dimension reduction in AEs plays a key role to enhance the performance of DNNs as important as sparsity in traditional imaging and image processing algorithms.

\begin{figure}[htbp]
\centering
\begin{minipage}[t]{0.6\textwidth}
\centering
\includegraphics[width=7cm]{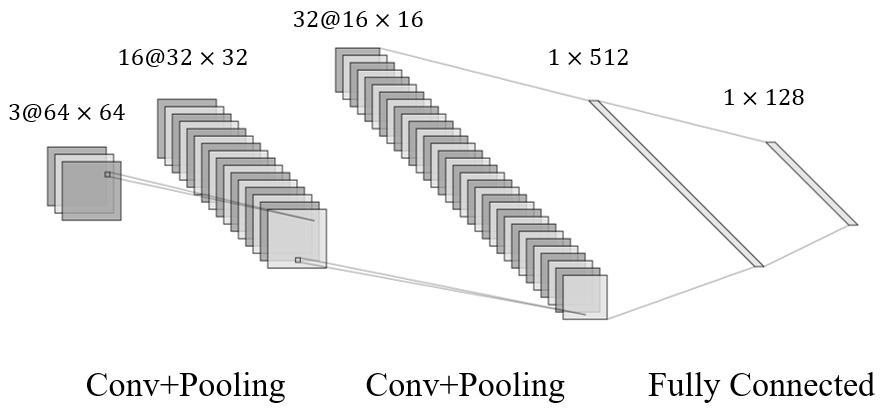}
\caption{Typical encoding network in convolutional autoencoder.}
\label{fig:CNN}
\end{minipage}
\begin{minipage}[t]{0.39\textwidth}
\centering
\includegraphics[width=5cm]{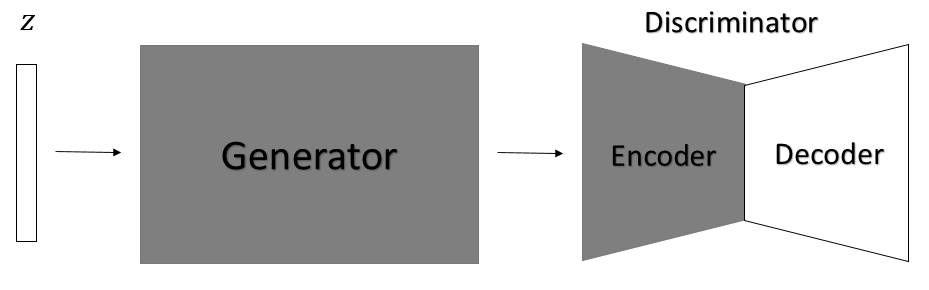}
\caption{BEGAN structure.}
\label{fig:BEGAN}
\end{minipage}
\end{figure}

Especially, deep convolutional encoders, $\EN(x;\theta_{EN})$ with parameters $\theta_{EN}$, can adaptively capture the low-dimensional structure of a source image $x$ through repeated applications of convolution, pooling, and nonlinear activation functions, and finally outputs a small feature vector $z$; and deep convolutional decoders, $\DE(z;\theta_{DE})$ with parameters $\theta_{DE}$, can efficiently reconstruct the image $x$ via a sequence of deconvolution, up-sampling, and nonlinear activation functions acting on $z$ (see Figure \ref{fig:CNN} for a visualization of the structure of an autoencoder).  Parameters in the encoder and decoder are jointly tuned such that the input and output of the autoencoder match well via the following optimization
\begin{equation}\label{eqn:AE}
\min_{\theta_{EN},\theta_{DE}} \mathbb{E}_{x\sim p_{data}(x)}\left[ \| \DE(\EN(x;\theta_{EN});\theta_{DE})-x\|_2^2\right], 
\end{equation}
where $p_{data}$ is the image data distribution. Armed with the powerful representation capacity of DNNs, deep convolutional autoencoders are capable of learning nonlinear transforms that create highly sparse or very low-dimensional representations of images from a certain distribution, while keeping the accuracy of image reconstruction. Due to the least square nature of \eqref{eqn:AE}, AEs penalizes pixel-wise error and hence prefers smooth images that lack fine details of the original image and avoids generative changes in image reconstruction.

Another stream of deep learning approaches (\cite{WB2017,BJPD2017,RCTAMM2017,YW2017,OVMDJ2018,CCCY2018}) is based on generative models, e.g., the generative adversarial network (GAN) (\cite{G2014}) and its variants (\cite{RMC2015,ACB2017,ZML2016,BSM2017}). A GAN consists of a generator $\G$ and a discriminator $\D$ that are trained through an adversarial procedure (see Figure \ref{fig:BEGAN} for a visualization of the structure of a GAN). The generator, $\G$, takes a random vector $z$ from a given distribution $p_z$ and outputs a synthetic sample $\G(z;\theta_G)$, where $\theta_G$ refers to the corresponding parameters of $\G$. The discriminator, $\D$ with parameters $\theta_D$, takes an input $x$ and outputs a value $\D(x;\theta_D)\in [0,1]$ denoting the probability of the input following the data distribution $p_{data}$. $\G$ is trained to generate samples to fool $\D$ into thinking that the generated sample is real, and $\D$ is learned to distinguish between real samples from the data distribution versus synthetic fake data from $\G$ via an adversarial game below:
\begin{equation}\label{eqn:GAN}
\min_{\theta_G}\max_{\theta_D} V(\theta_D,\theta_G) = \mathbb{E}_{x\sim p_{data}(x)}[\log \D(x;\theta_D)]+\mathbb{E}_{z\sim p_{z}(z)}[\log (1-\D(\G(z;\theta_G);\theta_D))].
\end{equation}
The discriminator $\D$ in the adversarial learning above can be applied to enhance image quality as in (\cite{WB2017}). The generator $\G$ can also be applied as a tool for data augmentation (\cite{Bowles2018,HLCWHL2018}) or as an inverse operator that returns the desired image from compressed measurements in compressed sensing (\cite{BJPD2017}). However, solving the optimization in \eqref{eqn:GAN} or its variants is challenging and the solution might not be stable, e.g., the generator might create images with unreasonable global content, although fine details of image content are better than those generated by AEs. 

In this paper, we introduce a novel generative encoder (GE) model that takes advantage of AEs and GANs simultaneously for generative imaging and image processing with applications in compressed sensing and imaging, image compression, denoising, inpainting, deblurring, and super-resolution. The GE model consists of a pre-training phase and a solving phase. In the pre-training phase, we separately train a GAN and an AE. The generator $\G$ in GAN can capture the data distribution of a given image set and hence can generate training data to enhance the encoder $\EN$ of the AE such that $\EN$ can better compress images following the target data distribution. In the solving phase, given a noisy image $x=\mathcal{P}(x^*)$, where $x^*$ is the target unknown image, $\mathcal{P}$ is an operator adding an addictive, or multiplicative, or convolutional noise, or equivalently given such an image $x$ in the compressed domain, i.e., given $m=\EN(x)$, we solve the optimization problem
 \begin{equation}\label{eqn:GE}
 z^*=\underset{z}{\mathrm{argmin}} \|\EN(S(\G(z)))-m\|_2^2+\lambda\|z\|_2^2
 \end{equation}
to recover the image $x^*$ in a generative way via $\hat{x}:=\G(z^*)\approx x^*$, where $S$ is a down-sample or an up-sample operator to balance the dimension of the output of $\G$ and the input of $\EN$, and $\lambda>0$ is a hyperparameter. The GE model unifies the generative capacity of GANs and the stability of AEs in an optimization framework in \eqref{eqn:GE} instead of stacking GANs and AEs into a single network or combining their loss functions into one as in existing literature.

The novelty and the advantages of the proposed GE model can be summarized as follows.
\begin{enumerate}
\item Different to existing methods, the training of GAN and AE in the GE model is separate to reduce the competition of GAN and AE to maximize the generalization capacity of GAN and the dimension reduction ability of AE. Numerical results will show that GE outperforms traditional methods, end-to-end convolutional DNN models, and previous GAN-based models achieving a better compression ratio. 
\item The training data of AE is augmented by the generator of GAN such that the dimension reduction of AE can adapt to the target data distribution, making the AE more compatible with GAN and increasing the generalization flexibility of AE.
\item We only unify the most attractive parts of GAN and AE, i.e., the generator of GAN for generalization and the encoder of AE for data-driven compression, which is different to existing works that combine all components of GAN and AE together in a single network.
\item Instead of creating an end-to-end neural network, a regularized least square problem \eqref{eqn:GE} is proposed to search for the best reconstruction that fits data measurements in the compressed domain. On the one hand, the reconstruction has been stabilized via reducing the search domain from the image domain to the compressed domain, e.g., the encoder can rule out extreme results generalized by the generator. On the other hand, the reconstruction still inherits the generalization capacity to enhance the detailed reconstruction which is often lost in traditional AEs.
\item The GE model is a general framework with various applications such as denoising, deblurring, super-resolution, inpainting, compressed sensing, etc. Numerical experiments show that the proposed model can generally produce competitive or better outcomes compared to existing methods.
\end{enumerate}

\section{Applications}
Let's briefly introduce the applications of the GE model in this paper.

\textbf{Compressed Sensing} 
Given a measurement vector $y = Ax^*  + \epsilon$, where $A$ is a sensing matrix satisfying the restricted isometry property (RIP), $\epsilon$ is a noise vector, the compressed sensing problem seeks to recover $x^*$. If $x^*$ is sparse, the recovery via an $\ell^1$-penalized least square problem is guaranteed by the compressed sensing theory in (\cite{CRT2006,D2006}). The RIP condition is satisfied when $A$ is a random Gaussian matrix, and natural images are generally sparse after an appropriate transform, e.g., wavelet transforms. Therefore, compressed sensing has been a successful tool in imaging science.

Recently, pioneer works including (\cite{BJPD2017}) have explored the application of generative models to improve traditional compressed sensing algorithms. The main idea is to apply the generator $\G$ to generate a synthesized image $\G(z)$ for $z$ in a compressed space. Bora et al. proposed 
\begin{equation}\label{eqn:CS}
z^*=\arg\min_{z} ||A \G(z)-y||_2^2+\lambda\|z\|_2^2
\end{equation}
to find the reconstruction image $\G(z^*)\approx x^*$. They also provided a theoretical guarantee on the relief of sparsity condition on the compressed space and thus serves as a nice benchmark for our model.

\textbf{Denoising and Inpainting}
Denoising and inpainting have the same problem statement in mathematics. Given a measurement $y=x^*+\epsilon$ or $y=x^*\circ \epsilon$, where $\epsilon$ is a certain random or structured noise and $\circ$ represents the Hadamard product, denoising and inpainting seek to recover $x^*$ from $y$. 

Traditional denoising or inpainting techniques generally rely on the sparsity of $x^*$ after an approximate transform in a certain metric, e.g., KSVD (\cite{ksvd}), GSR (\cite{ZZG2014}), BM3D (\cite{D2009}), NLM (\cite{BCM2005}), and total variation (TV) regularization (\cite{G2012}). Deep learning approaches have become more popular than traditional methods recently, e.g., deep autoencode methods (\cite{V2010,P2016,yu2018}), especially when the hidden information of noisy or damaged images is not visually obviously.

\textbf{Deblurring}
Blurring an image is commonly modeled as the convolution of a point-spread function over an original sharp image. Deblurring aims to reverse this process. Mathematically speaking, given a measurement $y=x * h$, where $h$ is an unknown convolution kernel function and $*$ represents the convolution operator. Deblurring seeks to recover $x$ with certain assumptions on $h$ and $x$ to ease the ill-posedness. Sparse coding (\cite{Dong2011}) and kernel estimation (\cite{Xu2010}) are effective methods for image deblurring. Deep convolutional networks have also been applied to this problem recently (\cite{YS2016}), especially with the help of GAN (\cite{BSM2017}).

\textbf{Super-resolution}
Super-resolution imaging aims at generating high-resolution images from low-resolution ones. For example, given a measurement $y=S(x^*)$, where $S$ is a down-sampling operator, or a convolution operator with a convolution kernel function decaying quickly in the Fourier domain. Super-resolution imaging seeks to recover $x^*$ with certain assumptions on $S$ and $x^*$ to ease the ill-posedness. Traditionally, Bicubic interpolation and sparse-coding (\cite{Dong2011}) are popular tools to increase image resolution. Recently, deep convolutional neural networks have also been applied to solve this problem with great success  (\cite{DeepPrior,Ledig2017}).

\section{Implementation of the Proposed Framework}
\label{sec:Proposed Framework}

\begin{figure}[h]
    \centering
    \includegraphics[width=0.75\textwidth]{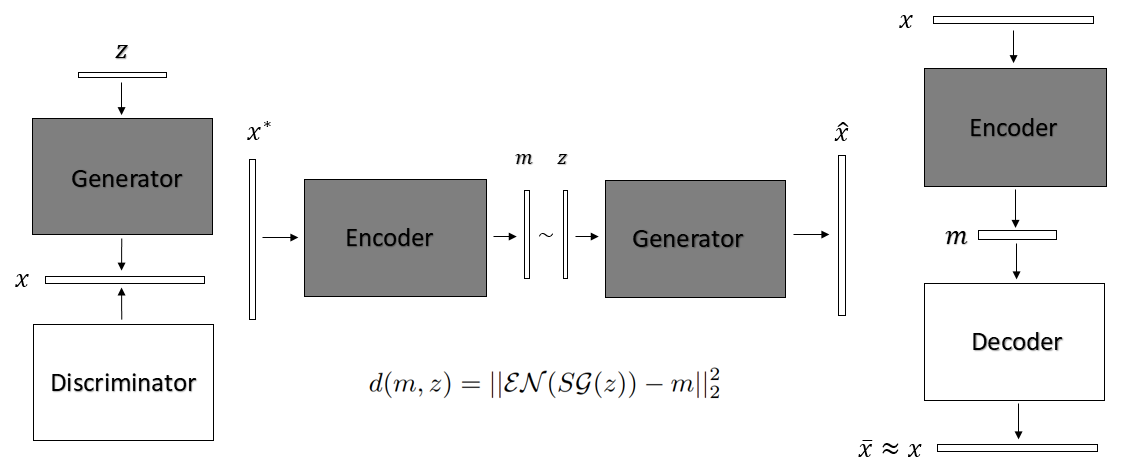}
    \caption{GE: Generative Encoder. Left: pre-trained generative adversarial network; middle: GE sensing and reconstructing flow; right: pre-trained convolutional autoencoder}
    \label{fig:EG}
\end{figure}

Let's introduce the detailed implementation of the GE model in this section. In this paper, we apply the BEGAN (\cite{BSM2017}) and the convolutional AE in the implementation. In fact, our framework is  broadly applicable to various GANs and AEs for the customized settings. The overall training procedure of the GE model is summarized in Algorithm \ref{alg:GE} and visualized in Figure \ref{fig:EG}.

\begin{algorithm}
\label{alg:GE}
\caption{GE: generative encoder model.}
\begin{enumerate}
  \item Pre-train a generator using any GAN.
  \item Pre-train a convolutional AE using both real and fake images (generated by GAN).
  \item Take the generator $\G$ of GAN and the encoder $\EN$ of AE to form the generative encoder.
  \item Given the measurement $m=\EN(x^*)$. Find
  $z^*=\arg\min_{z}||\EN(S(\G(z)))-m||_2 ^2+\lambda\|z\|_2^2$ and return $\hat{x}=\G(z^*)$.
\end{enumerate}
\end{algorithm}

The key idea of the GE model is to combine the generator $\G$ and the encoder $\EN$ from a GAN and an AE model, respectively, to take the advantages of these two models via a new optimization framework in \eqref{eqn:GE}. The term $d(m,z)= ||\EN(S(\G(z)))-m||_2^2$ itself could serve as the loss function of the GE model. We introduce the $\ell^2$ term to regularize the highly nonconvex function $d(m,z)$. Besides, looking for a solution with a small $\ell^2$-norm also agrees with the GAN model that $z\sim \mathcal{N}(0,I)$, i.e., $z$ has a higher probability to have a smaller magnitude. An interesting extension would be replacing the $\ell^2$ regularization with a DNN for a data-driven regularization. This is left as a future work.

\subsection{Compressed Sensing}
\label{subsec:reconstruction}

Suppose the original image is $x^*$ and we are able to design a sensor $\EN$. Given the measurement $m=\EN(x^*)$, in order to find the reconstruction $\hat{x}$ that is as close to $x^*$ as possible, we find $z$ in the latent space such that the compressed measurement of its generated image, i.e., $\EN(\G(z))$, is as close to the measurement of the real image $\EN(x^*)$ as possible. Therefore, we solve 
$$z^*=\arg\min_{z} ||\EN(\G(z))-m||_2^2 +\lambda\|z\|_2^2$$
to identify $\hat{x}:=\G(z^*)\approx x^*$ as the reconstruction.


\subsection{Denoising, Deblurring, Super-Resolution \& Inpainting}
\label{subsec:Denoising, deblurring, super-resolution inpainting}
The solution of the denoising, deblurring, super-resolution, and inpainting can be obtained by solving
\begin{equation}\label{eqn:de}
z^*=\arg\min_{z} ||\EN(S\G(z))-\EN(x^\dag)||_2^2 +\lambda\|z\|_2^2,
\end{equation}
where $x^\dag$ is the given noisy image, or the image with missing pixels, or the blurry image, or the image of low-resolution, constructed from an unknown target image $x^*$. Then the reconstructed image is set as $\hat{x}=\G(z^*)\approx x^*$. In \eqref{eqn:de}, $S$ is an adjustment operator, which is an identity for denoising and deblurring, a masking operator for inpainting, and a dimension adjustment operator for super-resolution. 

\section{Training Details \& Empirical Results}
\label{sec:Training Details & Empirical Results}

In this section, we describe training details and present experimental results to support the GE model. 

\subsection{Training Details}
\label{subsec:training details}
\textbf{Training Data:} We use the CelebA dataset (\cite{liu2015faceattributes}) containing more than $200,000$ celebrity images cropped to sizes of $64\times64\times3$ and $128\times128\times3$, respectively. 

\textbf{GAN:} The BEGAN is trained with the data sets above and the corresponding generator $\G$ is adopted in our GE model. The discriminator of the BEGAN is, in fact, an AE (See Figure \ref{fig:BEGAN} for the structure of the BEGAN). For $64\times 64\times 3$ and $64$ for $128 \times 128 \times 3$ datasets, the input dimensions of $\G$ are $128$ and $64$, respectively; the numbers of convolutional layers of the encoder in BEGAN are $12$ and $15$, respectively; the numbers of convolutional layers of $\G$ are $9$ and $11$, respectively, with one up-sampling every two convolutions. 

\textbf{AE:} We test two different types of AEs and the corresponding GEs are denoted as $GE0$ and $GE1$. In the first type, we use the encoding part of BEGAN’s discriminator as the encoder in GE. Due to the special encoding-decoding structure, the discriminator of BEGAN is trained to compress real and fake images in the same manner as an autoencoder. In the second type, we adopt convolutional autoencoders and train them separately with both real images and fake images generated by BEGAN. 
 
We use the following notation to denote the specific structure of the $\EN$ in $GE0$ and $GE1$. For example, $GE1\, (d=4,f=16,m=128)$ indicates that: 1) the encoder $\EN$ in our GE model has $4$ convolutional layers; 2) the output dimension of $\EN$ is $m= 128$; 3) in the $n$-th convolutional layer, the number of filters is $n\times16$ 
for $GE1$, and $\left \lceil{\frac{n}{3}}\right \rceil\times16$ for $GE0$.

\textbf{Optimization in GE:} ADAM optimizer (\cite{kingma2014method}) is used with a learning rate $0.1$ to solve \eqref{eqn:GE}. For images of size $64\times64\times3$, we use $500$ iterations with $2$ random starts to find the optimal reconstruction; for larger images, $700$ iterations with $2$ random starts are used.

\textbf{Baseline Methods:} We compare GE with two baseline methods under the same conditions: 

\begin{itemize}
\item \emph{Lasso}: Let $x^*$ be an unknown image and it has a sparse representation in a dictionary $\Psi$. Given $b=A\Psi^\dagger x^*$, where $\dagger$ denotes the pesudo-inverse and $A$ is a random Gaussian matrix, \emph{Lasso} solves the minimization problem:  $\beta^*=\arg\min_\beta||A\beta-b||_2^2+\alpha||\beta||_1$ to estimate $x^*$ by $\hat{x}:=\Psi\beta^*$. We choose $\alpha=0.1$ and $\Psi$ as an overcomplete discrete cosine dictionary for all experiments.\\

\item \emph{GA}: This benchmark uses the model in \eqref{eqn:CS} proposed by (\cite{BJPD2017}). Instead of DCGAN, which is used in the original work, we use BEGAN’s generator to make the comparison consistent since the GE model uses BEGAN. 
\end{itemize}

\subsection{Experimental Results}

\begin{figure}[h]
    \centering
    \includegraphics[width=0.85\textwidth]{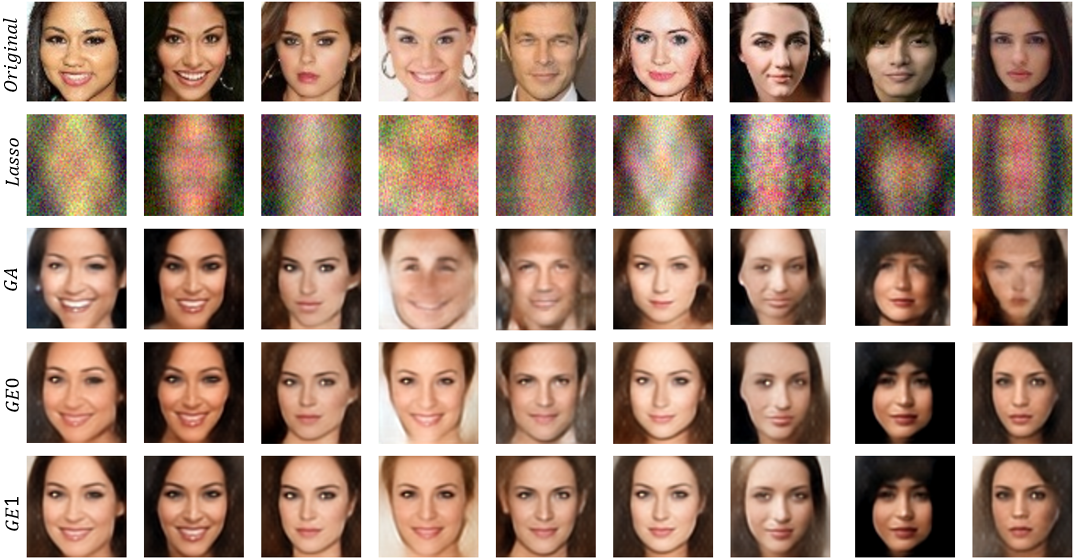}
    \caption{Reconstruction results on CelebA dataset $64\times64\times3$ images with $128$ measurements. $1^{st}$ row: original image; $2^{nd}$ row: $Lasso$ ($m=500$); $3^{rd}$ row: $GA$ ; $4^{th}$ row: $GE0\, (d=12,f=64,m=128)$; $5^{th}$ row: $GE1\, (d=4,f=16,m=128)$ }
    \label{fig:GE0}
\end{figure}

\textbf{Reconstruction - $64\times64\times3$ Images}
Figure \ref{fig:GE0} presents the comparison of $GE0$, $GE1$, and $GA$ with a measurement size $m = 128$ for $64\times64\times3$ images. The measurement size for $Lasso$ is set to $500$ because $Lasso$ totally fails when $m = 128$. 

According to visual inspection, GE model is more stable than $GA$ and $Lasso$ and the GE model can also preserve most fine details of faces as shown in Figure \ref{fig:GE0}. Actually, the reconstruction of $GA$ using DCGAN in (\cite{BJPD2017}) is comparable to our $GA$ reconstruction using BEGAN. Hence, the GE model outperforms $GA$ and $Lasso$.

Besides, two proposed frameworks, $GE0$ and $GE1$, have high compatibility and detail-retaining properties. Note that the network size of $GE0$ is much larger than that of $GE1$. Hence, we have shown that GE model is not sensitive to network size and is very efficient.

\begin{table}[h!]
\parbox{.45\linewidth}{
\centering
\begin{tabular}{cccc}
 \hline
   & Lasso & GA & GE0 \\
 \hline
 MSE & 0.024 & 0.015 & 0.009 \\  
 \hline\\
\end{tabular}
\caption{Reconstruction loss.}
\label{table:mse}
}
\hfill
\parbox{.45\linewidth}{
\centering
\begin{tabular}{cccc}
\hline
&Fake&Real&Ratio\\
 \hline
MSE & 0.00036 & 0.00928&0.039\\
\hline\\
\end{tabular}
\caption{MSE for real \& fake images.}
\label{table:errorsource}
}
\end{table}

To make a quantitative comparison, the sampled average of the mean square error (MSE) between restored and original images is computed for $GE0$ and $GA$ in Table \ref{table:mse}, which shows that $GE0$ outperforms $GA$ by $50\%$. In addition, we notice that the variation of the MSE of $GE$ generally is smaller $GA$, which also indicates that $GE$ is more stable than $GA$.

The reconstruction error comes from two sources: a systematic error caused by the compression and reconstruction mechanism, and a representation error caused by the gap between generated and real images.  The reconstruction loss of real and fake images is shown in Table \ref{table:errorsource}. The reconstruction for fake images generated by the generator only has one source of error: the systematic error. The result shows that more than $95\%$ of the current MSE comes from the gap between the generated space and real pictures. Therefore, if training data are better and the GAN design is better, the MSE can be further reduced and the GE model becomes better.

\begin{figure}[htbp]
\begin{minipage}[t]{0.6\textwidth}
\centering
\includegraphics[width=8cm]{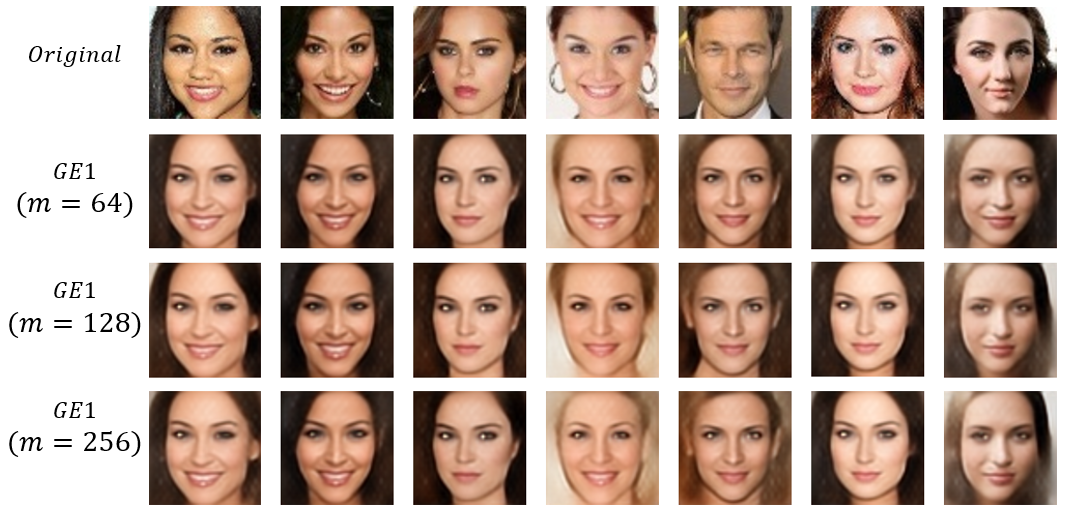}
\caption{Reconstruction with increasing measurement size. $1^{st}$ row: original image; $2^{nd}$ row: $GE1(d=4,f=16,m=64)$; $3^{rd}$ row: $GE1\, (d=4,f=16,m=128)$,$4^{th}$ row: $GE1\, (d=4,f=16,m=256)$}
\label{fig:measurement}
\end{minipage}
\hspace{0.5cm}
\begin{minipage}[t]{0.29\textwidth}
\centering
\includegraphics[width=5cm]{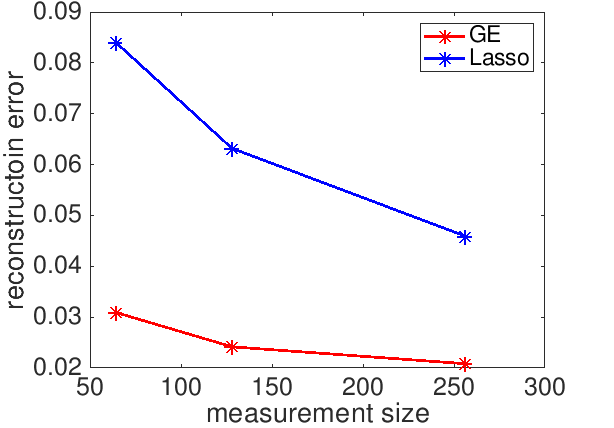}
\caption{GE1 reconstruction loss with different measurement sizes.}
\label{fig:MSE}
\end{minipage}
\end{figure}

Figure \ref{fig:measurement} visually demonstrates how measurement size affects the quality of reconstruction. Generally speaking, restoration with a larger measurement size has better quality, but the bound of the reconstruction error is still determined by the capacity of the generator. In addition, the reconstruction under measurement $m=64$ is still reasonable, although the restoration tends to be slightly distorted, which indicates that the GE model could be applied to significantly boost sensing efficiency. Figure \ref{fig:MSE} quantifies the influence of the measurement size on the MSE of image reconstruction. 

Finally, we would like to highlight that it is possible to enhance the encoder by sacrificing the decoder when training the autoencoder, as the decoder is not used anymore after training. By limiting the depth and width of the decoder, we can force the encoder to be more informative. Moreover, in different applications, customized encoders can be trained to cater for different data sets and challenges.

\begin{figure}[h]
    \centering
    \includegraphics[width=0.75\textwidth]{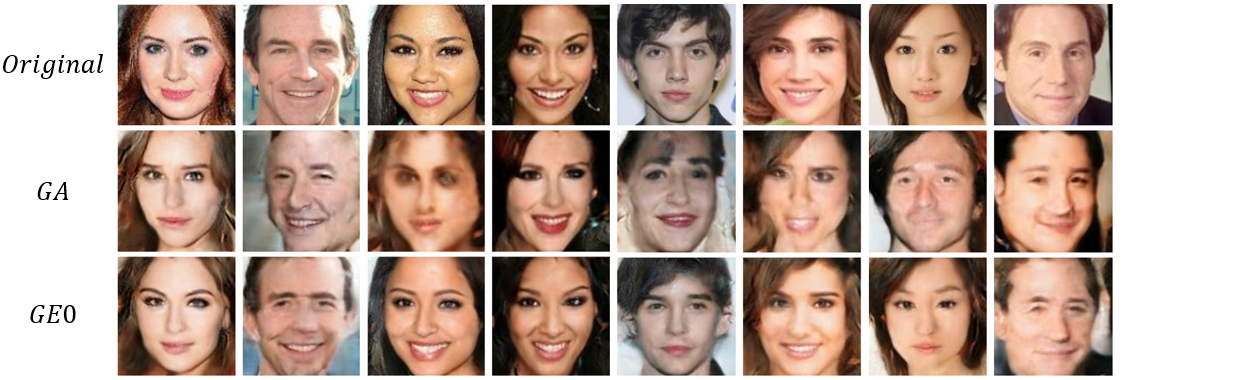}
    \caption{Reconstruction results on CelebA dataset $128\times128\times3$ images with $64$ measurements. $1^{st}$ row: original image; $2^{nd}$ row: $GA$; $3^{rd}$ row: $GE0\, (d=15,f=64,m=64)$}
    \label{fig:128}
\end{figure}

\textbf{Reconstruction - $128\times128\times3$ Images}  
Figure \ref{fig:128} shows a comparison between $GE0$ and $GA$ with a measurement size $m = 64$ for $128\times128\times3$ images with a measurement rate $\rho$ $0.0013$, which is very attractive for image compression. This measurement rate is mainly attributed to the quality of the pre-trained generator as well as the smoothness of large images. BEGAN itself has a very good generation quality for $128\times128\times3$ images and consequently, GE is able to deliver better results.

\begin{figure}[htbp]
\centering
\begin{minipage}[t]{0.49\textwidth}
\centering
\includegraphics[width=5cm]{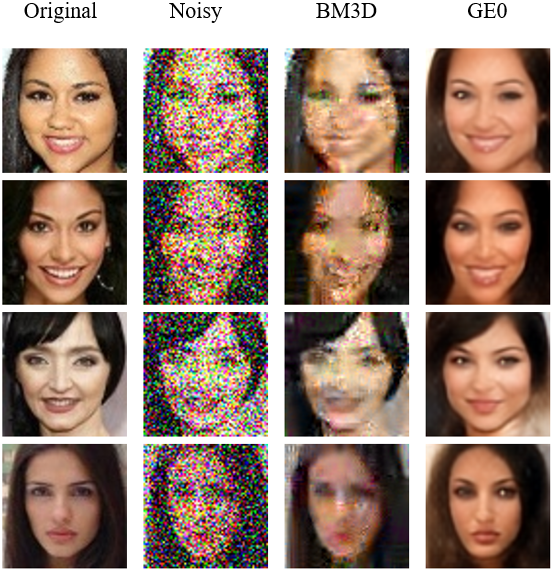}
\caption{Blind Denoising.}
\label{fig:denoising}
\end{minipage}
\begin{minipage}[t]{0.49\textwidth}
\centering
\includegraphics[width=5cm]{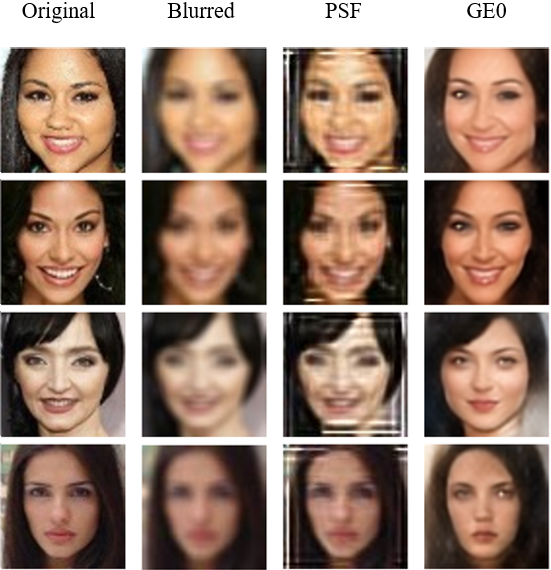}
\caption{Blind Deblurring.}
\label{fig:deblurring}
\end{minipage}

\end{figure}

\textbf{Denoising}
We add Gaussian random noise with standard deviation of $\sigma=0.4$ to pictures to generate noisy images. Suppose the noisy image is $x^\dag$, the GE model searches for a synthesized image that matches the input noisy image best in the compressed domain by solving
\begin{equation}\label{eqn:DN}
z^*=\arg\min_{z} ||\EN(S\G(z))-\EN(x^\dag)||_2^2 +\lambda\|z\|_2^2.
\end{equation}
The final reconstruction is given by $\hat{x}=\G(z^*)$. As Figure \ref{fig:denoising} reflects, benefited from the high compression rate due to the data-driven encoder of GE, GE is quite robust to noise, while BM3D cannot recover clear images when the GE model still works. 

\textbf{Deblurring}
To obtain a blurred image, rotationally symmetric Gaussian lowpass filter is applied to the clear images. To reverse the convolution process, blind deconvolution can be done under the same scheme as denoising. Suppose $x^\dag$ is the blurred image. The same algorithm as in \eqref{eqn:DN} is applied to reconstruct a clear image in the GE model. For comparison, a blind deconvolution based on point spread function (PSF) (\cite{Lam:00}) is listed together with GE in Figure \ref{fig:deblurring}. Compared with PSF, GE provides finer details on the resulting images. Again, the main source of error in the restoration by GE is the gap between generation and the original image. A well-trained GAN with sufficiently good training data can improve the performance of GE.

\begin{figure}[htbp]
\centering
\begin{minipage}[t]{0.49\textwidth}
\centering
\includegraphics[width=5cm]{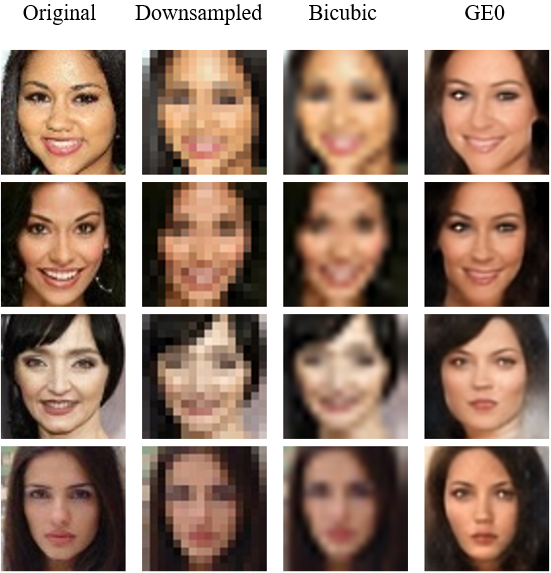}
\caption{Super-resolution.}
\label{fig:superres}
\end{minipage}
\begin{minipage}[t]{0.49\textwidth}
\centering
\includegraphics[width=5cm]{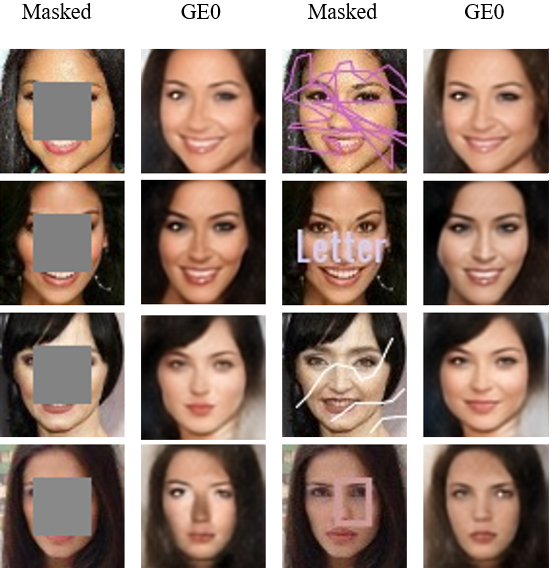}
\caption{Inpainting.}
\label{fig:inpaint}
\end{minipage}

\end{figure}

\textbf{Super-resolution}
We downsample original images from $64\times64\times3$ to $16\times16\times3$. The performance of Bicubic interpolation (\cite{8229459}) is included as a benchmark. Instead of inputting the downsampled image into GE directly, we use the image constructed by Bicubic interpolation as input $x^\dag$ in \eqref{eqn:DN}, which is equivalent to use the Bicubic interpolation as a preconditioner of the optimization in \eqref{eqn:DN}. Figure \ref{fig:superres} shows the results of the Bicubic interpolation and the GE model.

\textbf{Inpainting}
The first group of images is partially masked by a $30\times30$ block that sets the value in the missing area of the image $0$, and the second group is covered by hand-drawn lines and text. Suppose $\Omega$ is the mask area and $M$ is a binary matrix taking value $0$ at a position in $\Omega$ and value $1$ outside $\Omega$. Let $S$ be a masking operator such that $S(x)=x \circ M$, where $\circ$ is the Hadamard product. Again, assume $x^\dag$ is the corrupted input and reconstruct the inpainted images via \eqref{eqn:DN} with the masking operator $S$. Reconstructed images are shown in Figure \ref{fig:inpaint}.
\section{Conclusion}
In this paper, we have introduced the generative encoder (GE) model, an effective yet flexible framework that produces promising outcomes for generative imaging and image processing with broad applications including compressed sensing, denoising, deblurring, super-resolution, and inpainting. The GE model unifies GANs and AEs in an innovative manner that can maximize the generative capacity of GANs and the compression ability of AEs to stabilize image reconstruction, leading to promising numerical performance that outperforms several state-of-the-art algorithms. The code of this paper will be available in the authors' personal homepages.



\newpage 
\small
\bibliographystyle{abbrvnat}
\bibliography{refs}   




\end{document}